\documentclass[aip,jap,twocolumn,supercriptaddress,10pt]{revtex4-1}
\usepackage{amssymb}
\usepackage{amsfonts}
\usepackage{amsmath}
\usepackage[pdftex]{graphicx}
\usepackage{color}
\usepackage{dcolumn} 
\usepackage[normalem]{ulem}

\def\IR{\relax{\rm I\kern-.18 em R}}

\begin{document}
\title{Imaging flux distributions around superconductors: Geometrical susceptibility in the Meissner state}

\author{Mathieu N. Grisolia}
\affiliation{Laboratoire des Solides Irradi\'{e}s, CNRS-UMR 7642 \& CEA-DSM-IRAMIS, Ecole Polytechnique, F 91128 Palaiseau cedex, France}

\author{Antonio Bad\'{i}a--Maj\'os}
\affiliation{Departamento de F\'{\i}sica de la Materia Condensada--I.C.M.A., Universidad de Zaragoza--C.S.I.C., E-50018 Zaragoza, Spain}
 
\author{Cornelis Jacominus  van der Beek}
\affiliation{Laboratoire des Solides Irradi\'{e}s, CNRS-UMR 7642 \& CEA-DSM-IRAMIS, Ecole Polytechnique, F 91128 Palaiseau cedex, France}
\date{\today}

\begin{abstract}
Experiment and analytical calculations show that the demagnetizing field of a superconductor is a sensitive probe of  quantities otherwise difficult to measure, such as the sample-probe distance in flux-density imaging experiments, and the field of first flux penetration $H_{p}$. In particular, the ratio of the maximum field measured above the superconductor edge and the applied field can be determined unambiguously so as to define  a linear ``geometric'' susceptibility. The evolution of this susceptibility  with field depends on the regime of flux penetration, and can be used as a means to  determine  $H_{p}$ and the effect of a parallel field component in magneto-optical imaging experiments.

%High resolution magneto-optic experiments of superconducting crystals have been analyzed in the regime of low magnetic fields. Based on analytical calculations, we introduce new methods for determining quantities relevant to the measurements, viz.: the observation height over the superconductor and the spatially resolved first penetration field.  Revisiting Brandt-Mikitik's theory of the Meissner state in finite geometry, we show that the peaked structure of the flux line components above the sample behaves like a linear susceptibility. The scaling of this quantity as a function of the aspect ratio is presented and can be directly related to the observation height. Distortions of the ideal profiles caused by surface imperfections are also discussed. An accurate and fast determination of $H_{c1}$ is realized either in terms of the flux profile peaks as a function of the applied field, or from the analysis of the noise present in the experimental data. 

\end{abstract}
\maketitle
\section{introduction}

Magnetic imaging of superconductors\cite{Bending99} is widely used to extract parameters such as the superfluid density,\cite{LanLuan2010,LanLuan2011,Shapoval2011} the field of first flux penetration $H_{p}$,\cite{Lyard2004,Okazaki2009} vortex phase transition fields, \cite{Zeldov95II} and spatially resolved critical currents.\cite{Kees1} Present-day techniques generally measure the magnetic induction component $B_{\perp}$ perpendicular to the sample surface,  and include magneto-optical imaging (MOI),\cite{Kees1,Dorosinskii92,Johansen96,Jooss2002,Uehara2010} Hall array\cite{Lyard2004,Okazaki2009,Zeldov95II,Kees1,Hall-array} and scanning Hall-probe magnetometry,\cite{ScanningHall} scanning Superconducting Quantum Interference Device magnetometry (scanning SQUID),\cite{ScanningSQUID} and Magnetic Force Microscopy (MFM).\cite{LanLuan2010}  While efficient schemes have been devised to extract information concerning the distribution of current flow in the superconducting bulk from such experiments,\cite{Inversion-scheme,Johansen96}  important limitations remain.  Among these is the neglect of end-effects in thick samples, and a general lack of knowledge of the sample-probe distance, in many cases resulting from the manual positioning of the specimen. The sample-probe distance, which has an immediate bearing on the absolute values of current densities deduced from the  experiment, is usually guessed, or inferred from a fitting procedure of the measured flux profile. Related is the problem of accurately measuring $H_{p}$ in type-II superconductors. Since, in the Meissner state, magnetic flux wraps around the sample edge due to the demagnetizing effect,  a measurement at a given probe height will yield considerable ambiguity when it comes to determining whether vortex lines have penetrated the material or not, especially in the presence of strong flux pinning.  Moreover, the measured  $H_{p}$ and  Meissner slope $dB_{\perp}/d\mu_{0}H_{a}$ depend on the placement of the probe and on sample geometry ($H_{a}$ is the applied magnetic field and  $\mu_{0} = 4\pi \times 10^{-7}$ Hm$^{-1}$). The  observation distance above the surface results in measured flux profiles that are rarely in  agreement with model calculations,\cite{ref2,ref3} particularly when it comes to  the field distribution near the sample edge, a situation that complicates the reliable extraction of superconducting parameters.

Here, we show that the situation can also be put to one's advantage. Namely, when imaging the flux distribution around a superconductor in the Meissner state,  the London penetration depth $\lambda_{L}(T)$ can generally be neglected. Thus, the demagnetizing field, and, specifically, the maximum value $B_{\perp}^{peak}$ at the sample edge, depends on the sample geometry, its aspect ratio, and on the distance from the surface, but not on any parameters characterizing the superconducting state. Since the sample geometry is known, measurement of the demagnetizing field peak grants access to the distance of the probe above the sample surface.  Below, the dependence of $B_{\perp}^{peak}$ on  $H_{a}$ is used to define a linear ``geometric susceptibility'' $\chi_{\rm g}$. A plot is provided  that allows one to simply read off the probe-to-sample distance using the experimentally determined $\chi_{\rm g}$ for a specimen of given aspect ratio. Also, the field dependence of  $\chi_{\rm g}$ reflects whether vortices have penetrated the material or not. One can thus  determine $H_{p}$ by a measurement of the flux density at a point above the superconductor perimeter. Finally,  $\chi_{\rm g}$ can also be used to estimate  the effect of the in--plane magnetic-field component  on the measured luminous intensity in MOI experiments. 

In what follows, we first recapitulate on the typical experimental procedure for the imaging of flux density distributions. Even if, in the present case, the experiment concerns MOI of the iron-based superconductor Ba(Fe$_{0.925}$Co$_{0.075}$)$_{2}$As$_{2}$, basic results are independent of the method and the material. A  
theoretical framework for calculating flux distributions  around superconductors of realistic shape is introduced. Basically relying on analytical techniques, it presents less computational difficulties than previous work.  The comparison of measurements with calculations focuses on the relation between  $B_{\perp}^{peak}$ and $H_{a}$, which turns out to be a good alternative indicator of $H_{\rm p}$. 

%%%%%%%%%%%% FIGURE I %%%%%%%%%%%%%%%%
%
\begin{figure}[t]
\includegraphics[width=8cm]{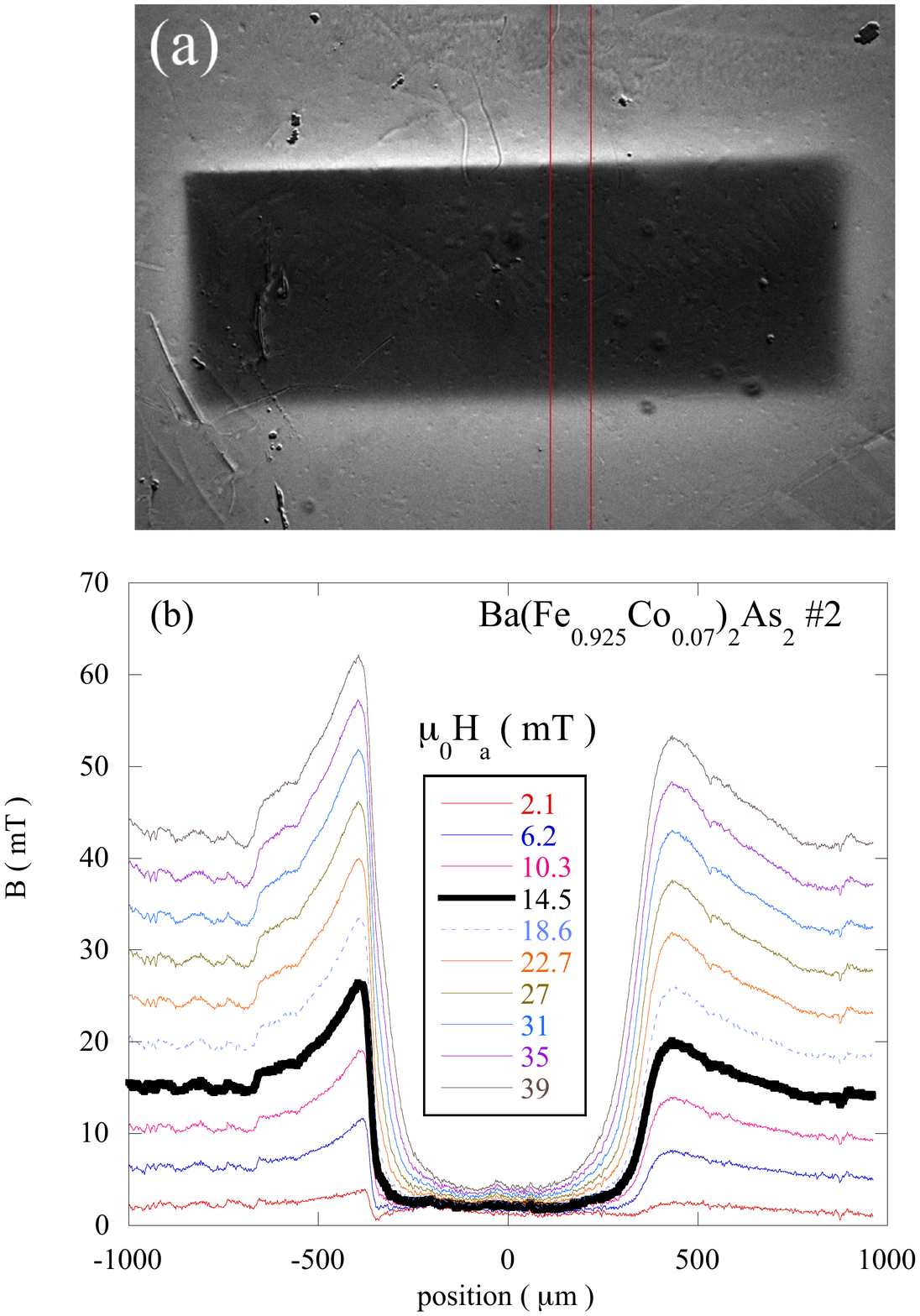}   
\caption{\label{Fig_1}(a) MOI of screening by  Ba(Fe$_{0.925}$Co$_{0.075}$)$_{2}$As$_{2}$ crystal \#~2, at $\mu_{0}H_{a} = 10$~mT, after zero-field cooling to 10~K. The rectangular outline of the crystal is clearly seen. The garnet has been purposely placed obliquely, so that the sample-to-garnet distance is smaller along the top edge than along the bottom edge. (b)  Profiles of the perpendicular flux density at successive applied fields, averaged over the strip between the two red lines in (a), after calibration of the luminous intensity. The heavy line indicates the first profile after flux penetration. The abscissa runs from the upper to the lower part of panel (a).}
%\end{center}
\label{fig:MOI}
\end{figure}  
%
%%%%%%%%%%%% FIGURE I %%%%%%%%%%%%%%%%

\section{Experimental Details}

Optimally--doped Ba(Fe$_{0.925}$Co$_{0.075}$)$_{2}$As$_2$ single crystals, with a critical temperature $T_{c} = 24.5$~K, were grown using the self-flux method.\cite{Rullier-Albenque2009} Rectangular samples were cut from different crystals using a W wire saw (wire diameter 20~$\mu$m) and 1~$\mu$m SiC grit suspended in mineral oil.  Sample \#~1 has length 994~$\mu$m,  width $2a = 571$~$\mu$m, and thickness $2b=32$~$\mu$m, while sample \#~2 has length 2200~$\mu$m, width $2a =  770$~$\mu$m and thickness $2b = 75$~$\mu$m.  Magnetic flux penetration into the selected samples was  visualized by the MOI method\cite{Dorosinskii92,Uehara2010} by placing a ferrimagnetic garnet indicator film with in-plane anisotropy directly on top of the sample. The garnet, of thickness 6~$\mu$m, was deposited by liquid-phase epitaxy on a 500 ~$\mu$m thick substrate, and is covered by a 100~nm-thick Al mirror layer. A non-zero $B_{\perp}$  induces an out-of-plane rotation of the garnet magnetization, and, thereby, a Faraday rotation of the polarization of the light traversing the garnet. The mirror layer reflects the impinging light, which is then observed using a polarized light microscope. Regions with non-zero induction show up as  bright when observed through an analyzer, nearly crossed with respect to the polarization direction of the incoming light.  Measurements of flux penetration were performed at different temperatures between 8 and 24 K.  

Fig.~\ref{Fig_1}(a) shows an example of the magnetic flux distribution around the Ba(Fe$_{0.925}$Co$_{0.075}$)$_{2}$As$_{2}$ crystal in the Meissner phase. The polarizer-analyzer pair was slightly uncrossed in order to obtain unambiguous results down to the lowest fields. Calibration of the luminous intensity with respect to $H_{a}$ allows one to convert the MO images into maps of $B_{\perp}({\mathbf r})$.\cite{Uehara2010}  Flux density profiles were determined parallel to the shorter sample dimension, close the center of the longer side. Previous measurements confirm that end effects induced by a finite sample length are irrelevant,\cite{Anna} as long  as this exceeds the width by a factor two.

\section{Physical modeling}
\label{section:calc}
We proceed by modeling the magnetic flux distribution around a rectangular superconducting parallelepiped, with the intent of achieving the least mathematical complication and the widest possible applicability.  The situation is considered in which a uniform magnetic field is applied perpendicularly to a long, ideally superconducting beam of rectangular cross section, considered infinite along the $z$-axis. The  magnetic flux density  ${\mathbf B}(x,y)$ is evaluated at a small distance above the surface. For very thin samples, the problem is quasi-one dimensional (quasi 1D); in this case, the inversion schemes of Refs.~\onlinecite{Jooss2002,Inversion-scheme,Johansen96} are satisfactory.  However, for samples of arbitrary  thickness $2b \parallel y$ and length $2c \parallel z$, sufficiently large with respect to the width $2a \parallel x$  (\em i.e. \rm double the width),  it is two-dimensional (2D).  This situation was previously  considered by Brandt and Mikitik.\cite{ref2}
The main findings of Ref.~\onlinecite{ref2} generalize Brandt's previous work for thin samples.\cite{ref3} 

In the case of a thin strip, the  cross-section of which corresponds to a line segment  $-a\leq x \leq a$, the Meissner surface current density is $J(x)=2 H_{a} x/\sqrt{a^{2}-x^{2}}$. Inserting this  into Biot-Savart's law, and integrating in the complex plane  ($x+i y \equiv r {\rm e}^{i\varphi}$), one gets the flux density map around the sample. In particular, we obtain
\begin{equation}
\label{eq_compact}
{\displaystyle [B_{x}(x,y),B_{y}(x,y)] =\frac{{\mu_0}H_{a}}{\sqrt{c}} \left[{\rm sin}(\alpha/2),{\rm cos}(\alpha/2)\right] . \quad}
\end{equation}
Here, we have defined $\alpha\;   \equiv    \tan^{-1} [\sin{2\varphi}/(r^2-\cos{2\varphi})]$
and $c\;  \equiv    \sqrt{1-2r^2\cos{2\varphi}+r^4}/r$, and give the distances in units of $a$.

A similar approach may be applied to long samples of rectangular cross section ($-a\leq u\leq a \; , -b\leq v\leq b$) based on the following expressions for the surface current density\cite{ref2}
\begin{eqnarray}
\label{eq_rectangle-a}
J(u, v = \pm b) & = & \frac{H_{a}s_u}{\sqrt{1-s_{u}^2}}
\\
J(u=\pm a, v) & = & \frac{\pm H_{a}\sqrt{1-m s_{v}^2}}{\sqrt{m(1-s_{v}^2)}}.
\label{eq_rectangle-b}
\end{eqnarray}
$s_{u}(u,m)$ and $s_{v}(v,m)$ are geometry dependent functions that may be calculated in terms of a parameter, $m$,\cite{ref4} that solely depends on the sample's aspect ratio $b/a$. The magnetic field around the beam can be obtained from Biot-Savart's law, by numerical integration over the four beam surfaces.\cite{ref5}
%%%%%%%%%%%% FIGURE II %%%%%%%%%%%%%%%%
%
\begin{figure}[t]
\includegraphics[width=8cm]{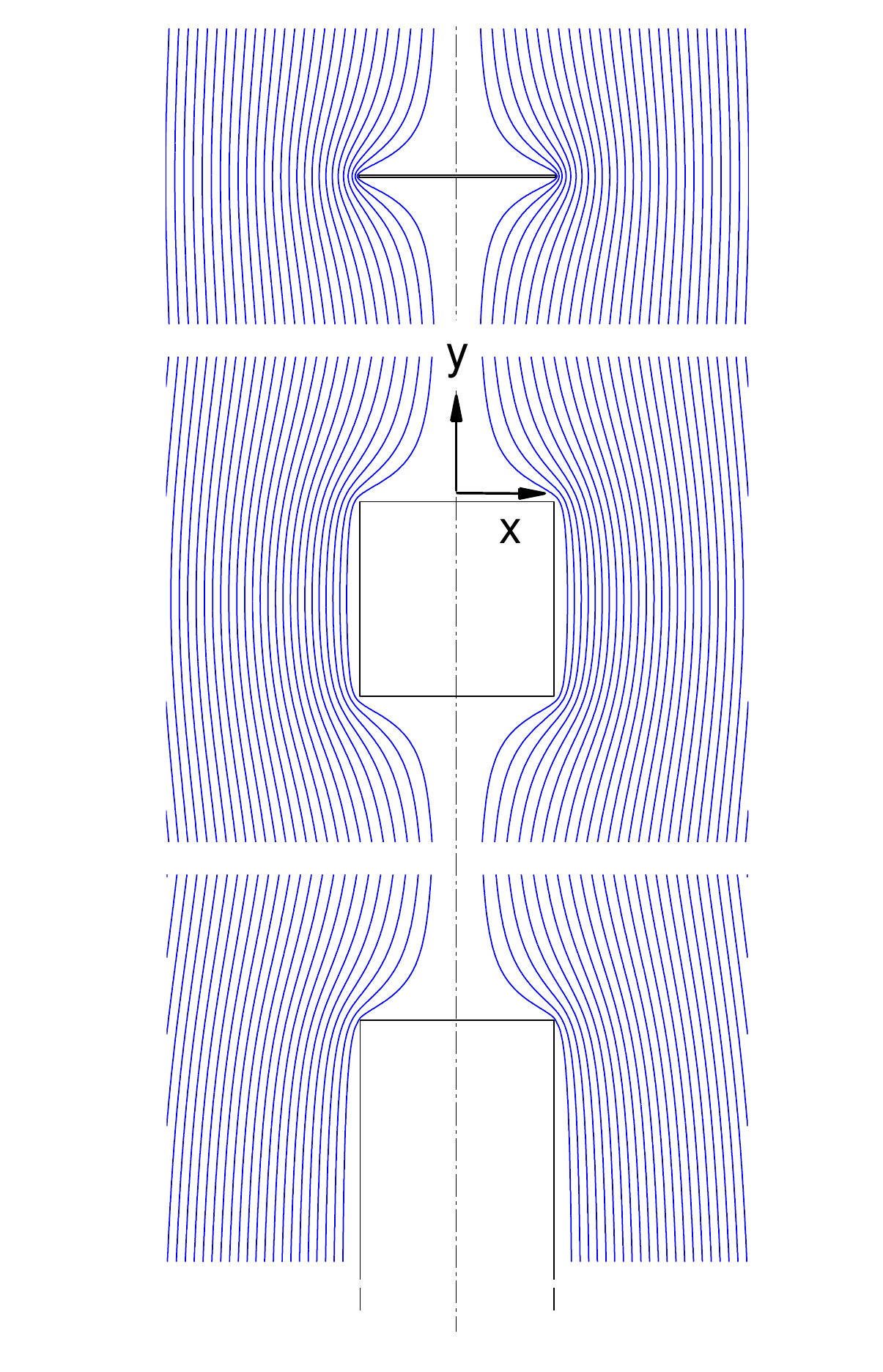}
\caption{\label{Fig_2}(Color online)  Theoretical magnetic field lines surrounding ideal superconducting beams of rectangular cross section, in the Meissner state. Results are shown for  aspect ratios $b/a = 0.001, 1$ and $10$ and are obtained from Eqs.(\ref{eq_rectangle-a}) and (\ref{eq_rectangle-b}).}
\end{figure}
%
%%%%%%%%%%%% FIGURE II %%%%%%%%%%%%%%%%

\begin{figure}[t]
\includegraphics[width=8cm]{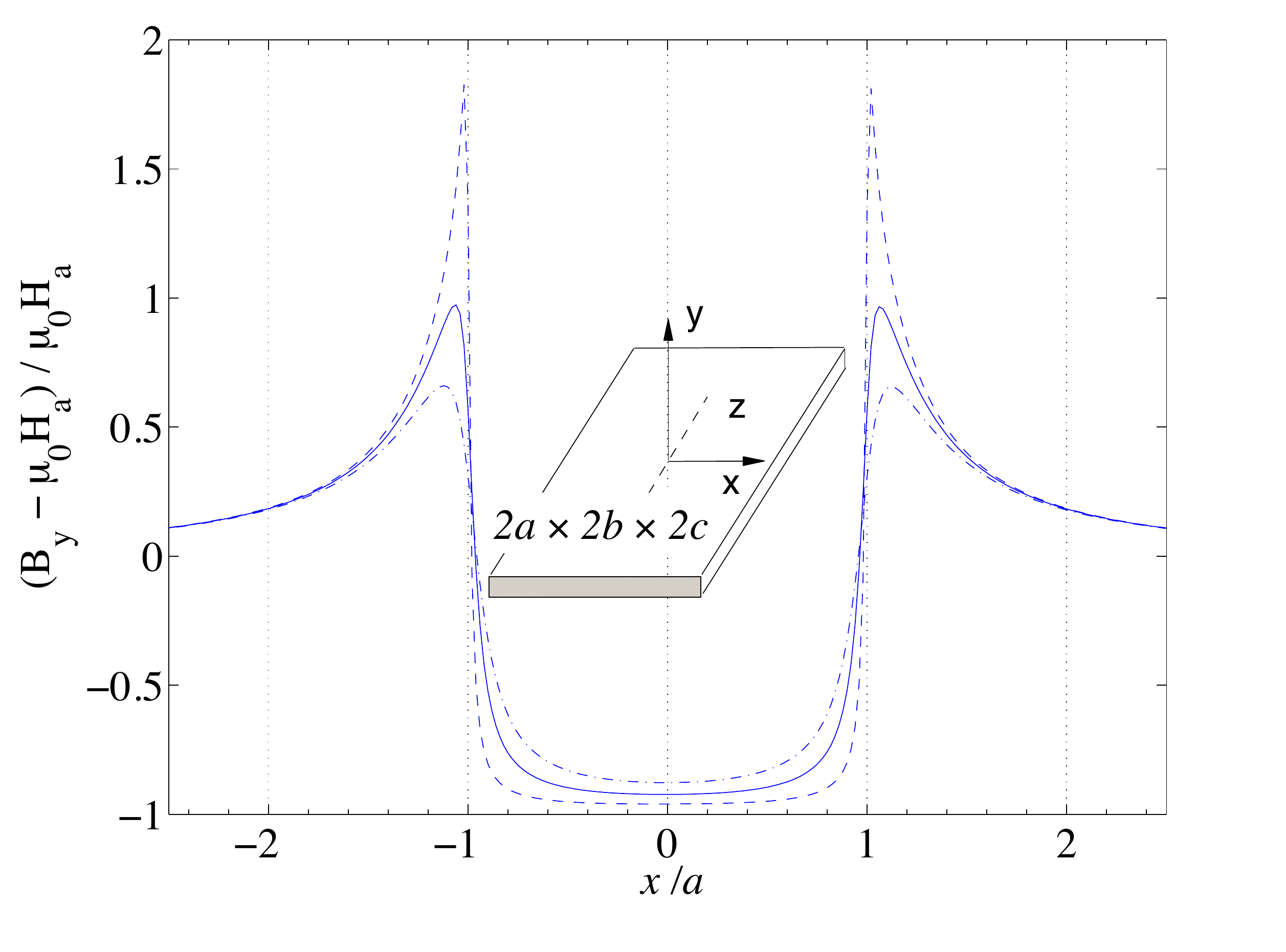}
\caption{\label{Fig_3}(Color online)  The distribution of  $B_{\perp}$ across the width of the superconducting beam, at three different heights $y$ above the surface, such as calculated for an aspect ratio $b/a = 0.1$. Different lines correspond to $y/a = 0.01$ (dashed), $0.05$ (continuous) and $0.1$ (dotted-dashed). In all cases, a uniform applied magnetic field  $(0,H_{a},0)$ is assumed.}
\end{figure}

\begin{figure}[b]
\includegraphics[width=8cm]{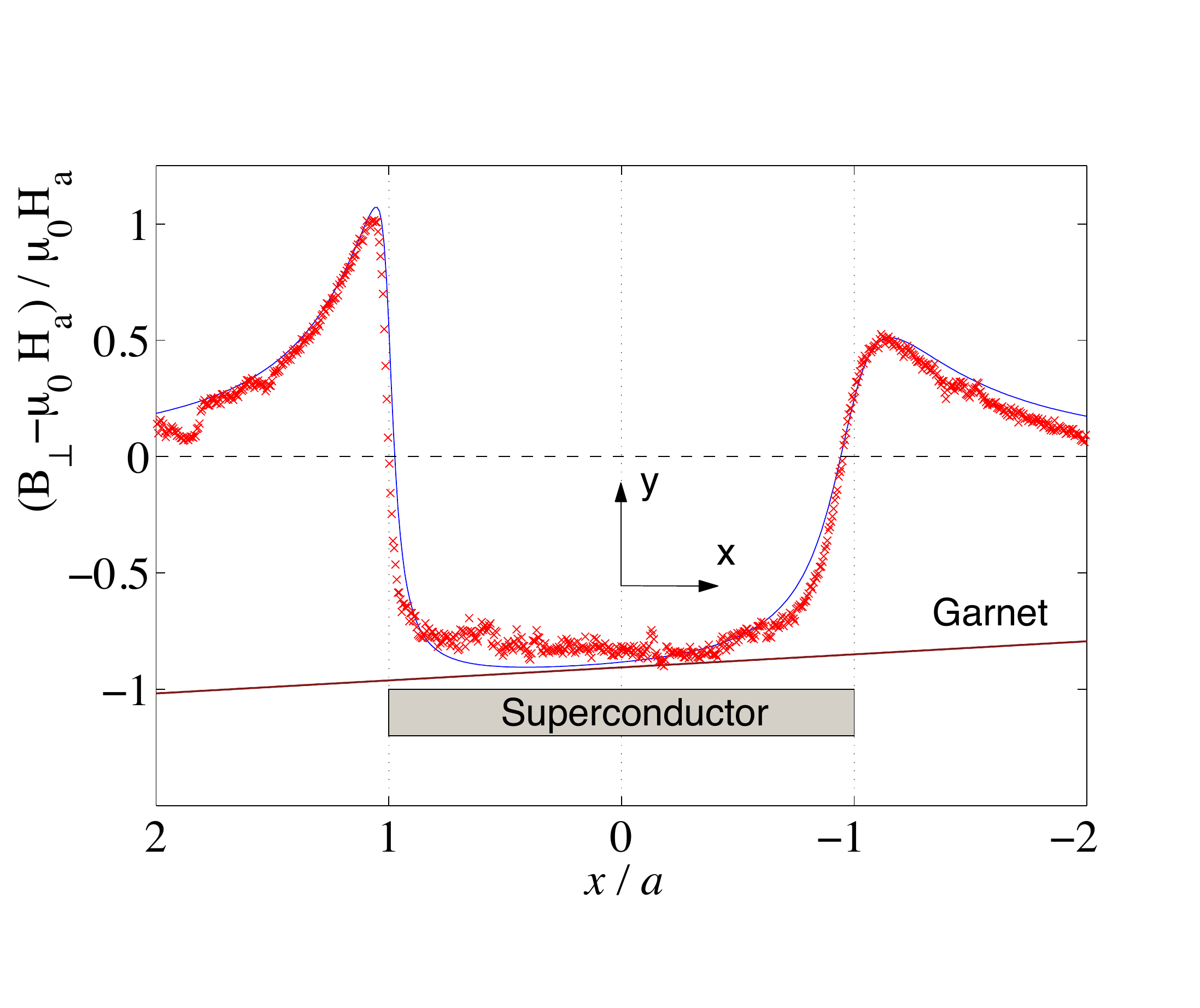}
\caption{\label{Fig_4}(Color online)   Measured and calculated flux density profiles when the MO garnet is placed obliquely over  sample \#~2. The crosses are the experimental points, while the drawn line denotes the calculation. The latter was carried out using the optimized distances  $y = 0.038 a$ (14~$\mu$m)   and  $y = 0.15 a$ (58~$\mu$m) above the left and right edges, respectively. In this case, the crystal of aspect ratio $b/a = 0.10$ was used. After correction for the in-plane field effect on the garnet magnetization (section~\protect\ref{section:in-plane}) these distances become 10 and 41~$\mu$m, respectively.}
\end{figure}

Fig.~\ref{Fig_2} displays the field lines calculated for such superconducting beams, in the Meissner state, with different values of the aspect ratio $b/a$. The  simulated flux density distribution at different heights above the sample such as this would be imaged, {\em e.g.}, with MOI, is shown in Fig.~\ref{Fig_3}. Note that the sharp bending of the field lines around the sample ridges produces the well-known $B_{\perp}(x)$--distribution, with sharp maxima due to the demagnetizing effect over each edge. The peak becomes progressively sharper as the observation height and/or sample thickness diminish. Physically, the condition $\nabla \times {\bf B}=0$ above the sample imposes $\partial_{y}B_{x}=\partial_{x}B_{y}$; thus, a more pronounced bending of the field lines (increasing $\partial_{y}B_{x}$) is accompanied by a growing value of the profile's slope $\partial_{x}B_{y}$. Also, whereas the sharpest peaks, measured in intimate contact with the superconductor, appear right above the edges, the maxima of the flatter peaks measured at larger height are located outwards. Finally, asymmetric profiles are found when the imaging device is oblique with respect to the superconductor surface, as in Fig.~\ref{Fig_4} -- note that in this case $B_{\perp} \neq B_{y}$. The plot shows our experimental data together with a least squares fit profile obtained by minimizing the difference between data and theory. The heights of the garnet above the left and right edges are used as  optimization parameters.

%%%%%%%%%%%% FIGURE III %%%%%%%%%%%%%%%%
%
%
%%%%%%%%%%%% FIGURE III %%%%%%%%%%%%%%%%

%%%%%%%%%%%% FIGURE IV %%%%%%%%%%%%%%%%
%
% 
%%%%%%%%%%%% FIGURE IV %%%%%%%%%%%%%%%%

%
%%%%%%%%%%%% FIGURE V %%%%%%%%%%%%%%%%
%
\begin{figure}[t]
\includegraphics[width=8cm]{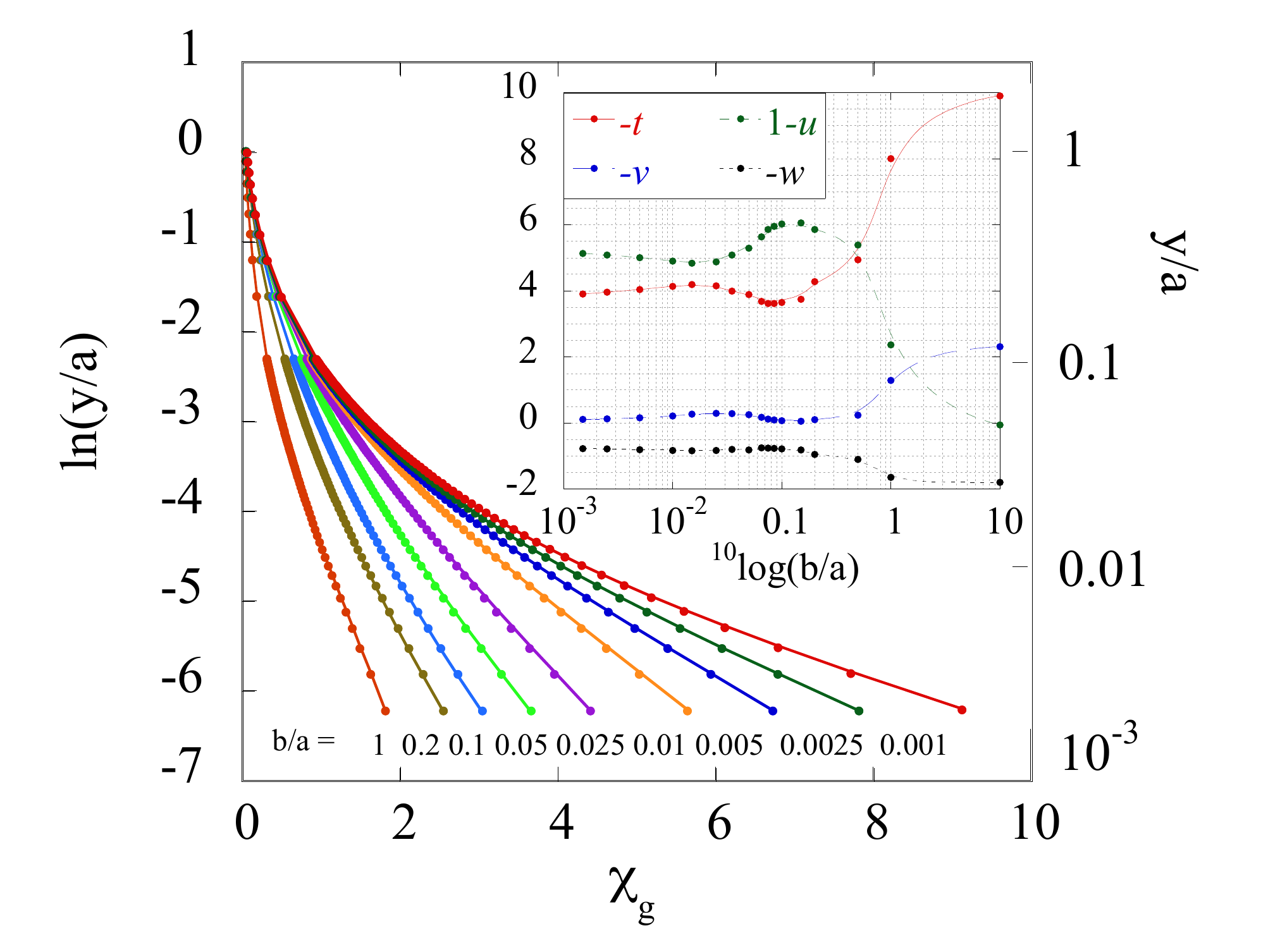}
\caption{\label{Fig_5}  Relation between the distance $y$ above the sample surface (in terms of the sample width $a$) and $\chi_{\rm g}$. 
%. The continuous lines denote the graphical inversion of the inset curves calculated using Eqs.~(\protect\ref{eq_rectangle-a}) and (\protect\ref{eq_rectangle-b}). Dots are result of the fit given by Eq.~(\protect\ref{eq_linfit_inv}). (Inset) The peak of the superconductor's magnetic field ($\chi_{\rm g}\equiv(B_{\perp}^{peak}-\mu_{0}H_{a})/\mu_{0}H_{a}$) in terms of the observation distance $y$ above the surface. 
The curves, from left to right, correspond to aspect ratios  $b/a = 1, 0.2, 0.1, 0.05, 0.025, 0.01, 0.005, 0.0025$, and 0.001. The continuous lines are obtained by numerical integration of Biot-Savart's law, combined with Eqs.~(\protect\ref{eq_rectangle-a}) and (\protect\ref{eq_rectangle-b}). Symbols correspond to the fit given by Eq.(\ref{eq_linfit_inv}). Inset: the aspect ratio-dependent parameters $t$, $u$, $v$, and $w$.}
\end{figure}
%
%%%%%%%%%%%% FIGURE V %%%%%%%%%%%%%%%%
%
%%%%%%%%%%% FIGURE VI %%%%%%%%%%%%%%%%
%
\begin{figure}[b]
\includegraphics[width=8cm]{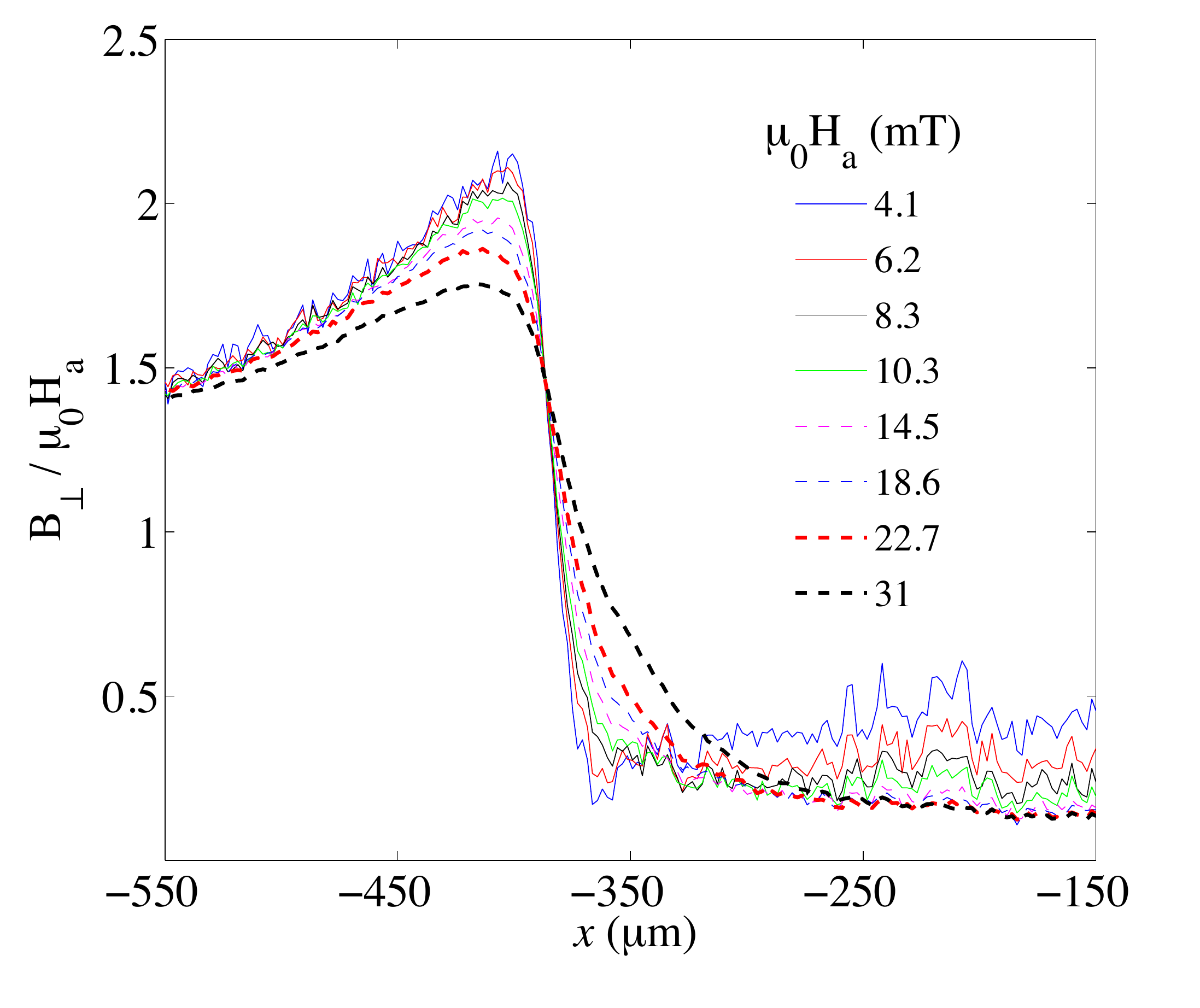}
 
\caption{\label{Fig_6} Renormalized flux profiles over   Ba(Fe$_{0.0925}$Co$_{0.075}$)$_{2}$As$_{2}$ crystal \#~2, for  $4.14 < \mu_{0}H_{a}< 30.95$~mT, at $T = 11$~K. %The inset shows the behavior of the profile noise as a function of the applied field. Noise is identified with the variance of the signal. We see that the noise is strongly reduced after a certain applied field, that matches the first penetration field defined before. This is indicated by the linear extrapolation of the variance to zero.
}
\end{figure}
%
%%%%%%%%%%%% FIGURE VI %%%%%%%%%%%%%%%%

\subsection{The peak susceptibility.}
To quantify the response of a given sample in a given experiment, one should consider the dependence of the peak value  $B_{\perp}^{peak}$ of the magnetic flux density profile on the distance above the sample edge.  Fig.~\ref{Fig_3} shows the behavior of the field contributed by the superconductor, in units of the applied field, {\em i.e.} $B^{s}_{y}/\mu_{0}H_{a} \equiv (B_{y}^{}-\mu_{0}H_{a})/\mu_{0}H_{a}$. With the superconductor in the Meissner state, this quantity depends only on the geometry of the sample and of the experimental arrangement, and is therefore independent of magnetic field. Thus, one can define  a linear {\em geometric susceptibility} $\chi_{\rm g} \equiv (dB_{\perp}^{\rm peak}/d\mu_{0}H_{a}) -1$. The fact that this is a purely geometrical quantity is clear from  Eqs.~(\ref{eq_compact}--\ref{eq_rectangle-b}). The choice of the field peak for the definition of $\chi_{\rm g}$ is preferable over that of more ambiguous features.

Fig.~\ref{Fig_5} shows the relation between  $\chi_{\rm g}$, the aspect ratio $ b/a$, and the observation height $y$. 
%An excellent fit of $\chi_{\rm g}$ is given by the expression 
%\begin{equation}
%\chi_{\rm g} \approx -{\rm erf}[\log({y}/{a})/{\lambda}] (\mu\, {\rm log}({{y}/{a}}) + \nu),
%\end{equation}
% with $\lambda ,\mu ,\nu$  aspect ratio-dependent parameters. Note the linearity for very small heights above  the sample surface. 
Indeed, the plot displays the more interesting inverse function $y(\chi_{\rm g})$, since this  allows one to obtain the observation height $y$ in terms of $\chi_{\rm g}$. A useful fit of $\ln({y}/{a})$ as a function of $\chi_{\rm g}$, with a relative quadratic error of $10^{-5}$, is given by
\begin{equation}
\label{eq_linfit_inv}
\displaystyle{\ln({y}/{a}) \approx t\, {\chi_{\rm g}^{{1}/{2}}} + v\, \chi_{\rm g}^{{3}/{2}}+ u  \chi_{\rm g}+w}.
\end{equation}
The aspect ratio-dependent parameters $t$, $u$, $v$ and $w$ are given in the inset of Fig.~\ref{Fig_5}.

%Eq.(\ref{eq_linfit_inv}) is a central result of this paper, and can be used as a tool for the analysis of magnetic induction maps above superconductors. Thus, in order
To determine the sample-to-probe distance, one should proceed in the following manner:
(i) determine the aspect ratio of the sample 
(ii) perform the measurement of the flux density distribution, ensuring a good accuracy, especially at low fields ($H_a  < H_p$); in MOI, this entails uncrossing the polarizer and analyzer by a small angle $\alpha$ during the measurement. 
(iii)  compute $\chi_{\rm g}$ from a linear fit of the low-field dependence of the maximum of $B_{\perp}(x)-\mu_{0}H_a$ (ideally, this coincides with the value of the peak itself), and
(iv) use Eq.(\ref{eq_linfit_inv}) and the graphical determination of the aspect-ratio dependent parameters (Fig.~\ref{Fig_5}) in order to determine $y$.  Note that the above analysis relies on the linearity of the response of the superconductor as function of the applied magnetic field $H_{a}$, and therefore  can be applied only for $H_a  < H_p$. 

%%%%%%%%%%%% FIGURE VII %%%%%%%%%%%%%%%%
%
\begin{figure}[t]
\includegraphics[width=8cm]{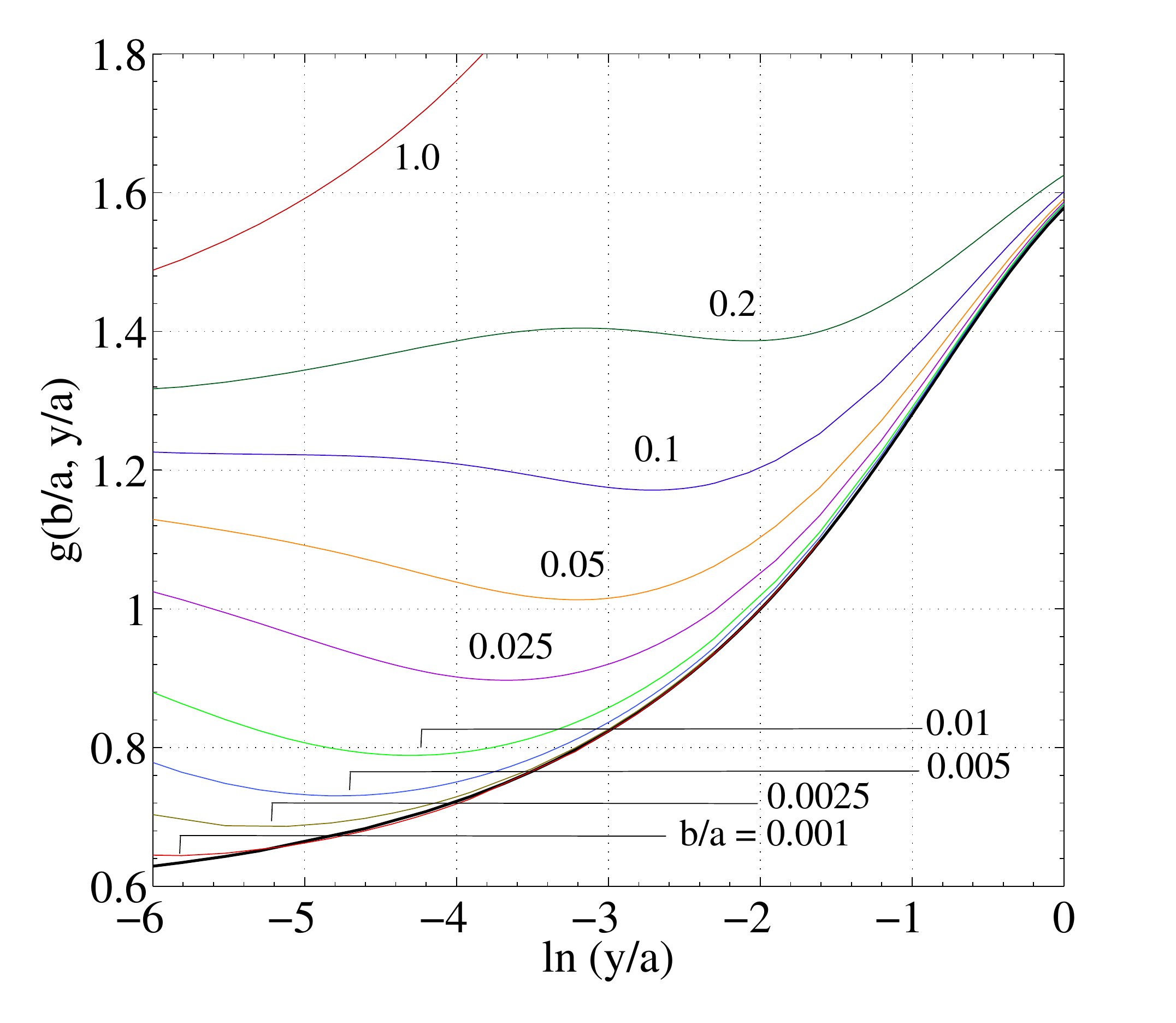}
\caption{\label{Fig_7} The function $g(y/a,b/a)$ relating the in--plane and perpendicular field components {\em i.e.}: $B_x$ and $B^{s}_{\perp}$ at  the position where $B^{s}_{\perp}$ is maximum above the superconducting beam (see text). It is plotted as function of the sample-to-probe distance $y/a$ for different aspect ratios $b/a$. Drawn lines correspond to the different values of $b/a$ considered in Fig.\ref{Fig_5} as labelled. The thick line comes from the application of Eq.(\ref{eq_compact}) for the thin sample limit.}
\end{figure}
%
%%%%%%%%%%%% END FIGURE VII %%%%%%%%%%%%%%%%
%
 \subsection{The field of first flux penetration}

On the contrary, a deviation from linearity can be used as a  criterion for determining $H_{p}$. The determination of $H_{p}$ by means of magnetic imaging is usually a somewhat time-consuming task, typically based on the detection of the minimum field that produces flux trapping in cyclic measurements.\cite{avraham} Also, the detection of the first vortices to enter the superconductor is clearly position-dependent. Our results suggest an alternative method. Below $H_{p}$, the magnetic flux is fully expelled from the sample, and the behavior of $\chi_{\rm g}$ is determined by the geometry of the experiment only. On the other hand, the evolution of $B_{\perp}^{\rm peak}$ beyond $H_{p}$ will not be linear in $H_{a}$ anymore, because it will reflect the flux pinning properties of the superconductor. This is explicitly shown in Fig.~\ref{Fig_6}. For fields lower than $H_{p}$, the demagnetization peak above the sample edge can be superposed by a simple rescaling by the value of the applied field. Beyond $H_{p}$, this scaling property is lost.

\subsection{Effect of an in--plane field component in MOI}
\label{section:in-plane}
The use of garnet indicators with in-plane anisotropy for the imaging of field distributions\cite{Dorosinskii92,Johansen96,Jooss2002} has the drawback that a magnetic-field component $B_{x}$ parallel to the indicator plane diminishes the Faraday rotation of the garnet magnetization. The magnitude of the effect increases as the screening current in the underlying superconductor increases, leading to a downward deviation from linearity of the $B^{peak}_{\perp}(H_{a})$--relation even in the Meissner phase, thereby complicating the determination of $\chi_{\rm g}$ and $H_{p}$. 

However, the linearity of the electromagnetic response in the Meissner state allows one to correct for the in--plane field effect in a relatively simple manner. The measured luminous intensity depends on the perpendicular field component $B_{\perp}$ as\cite{Johansen96} 
\begin{equation}
I = I_{0} \sin^2\left({\mathcal V} M_{s} \frac{B_{\perp}}{\sqrt{(B_{x}+B_{K})^{2}+B_{\perp}^{2}}} + \alpha\right), 
\label{eq:intensity}
\end{equation}
where $I_{0}$ is the impinging intensity, $M_{s}$ and $B_{K}$ are, respectively, the saturation magnetization and the anisotropy field of the garnet,  and $\mathcal V$ is a constant. Neglecting the influence of $B_{x}$ leads to the determination of an experimental 
\begin{equation}
B_{\perp}^{MOI} = \frac{B_{K} B_{\perp} }{B_{x} + B_{K}}
\label{eq:exp-real}
\end{equation}
rather than the real perpendicular field component $B_{\perp}$.\cite{note} Writing $B_{\perp} = B^{s}_{\perp} + \mu_{0}H_{a} = \mu_{0}H_{a}(\chi_{\rm g}+1) $ as the sum of the magnetic induction contributions coming from the superconductor and from the applied field, respectively, we can solve for $B^{s}_{\perp}$. Namely, not only is the non-zero in-plane field component  determined solely by the presence of the superconductor,  the linear response in the Meissner state implies that for a given value of $x$ it can be written as $B_{x} = g(b/a,y/a) B_{\perp}^{s}$. Here, we have calculated the proportionality constant $g(b/a,y/a)$ relating the in--plane and perpendicular field components above the position $x$ at which $B^{s}_{\perp}$ is maximum. Apparently, this quantity depends on aspect ratio and sample-probe distance, but, as long as $H_{a} < H_{p}$, not on the magnetic field.  Solving Eq.~\ref{eq:exp-real}, we obtain
%\begin{equation}
%\chi_{\rm g} = \frac{\chi_{\rm g}^{MOI}}{1 - g B_{\perp}^{MOI}/B_{K}}. 
%\label{eq:chi-exp-real}
%\end{equation}
%
\begin{equation}
\chi_{\rm g} = \frac{\chi_{\rm g}^{MOI}}{1 - g (1+\chi_{\rm g}^{MOI})B_{a}/B_{K}}. 
\label{eq:chi-exp-real}
\end{equation}
$\chi_{\rm g}^{MOI}$ is the apparent geometrical susceptibility such as determined from the MOI experiment. The function $g(b/a,y/a)$ has been evaluated as the ratio of the superconductor's contribution to the in--plane and perpendicular field components such as calculated in subsection~\ref{section:calc}, and is shown in Fig.~\ref{Fig_7} for the readers' reference. 
Once the sample aspect ratio and the anisotropy field of the garnet indicator are known, the effect of the in-plane field can be determined by estimating $g(b/a,y/a)$ from Fig.~\ref{Fig_7} using  $\ln(y/a)$  read from Fig.~\ref{Fig_5}, and calculating a refined geometric susceptibility from Eq.~(\ref{eq:chi-exp-real}).

%%%%%%%%%%%% FIGURE VIII %%%%%%%%%%%%%%%%
%
\begin{figure}[t]
\includegraphics[width=8cm]{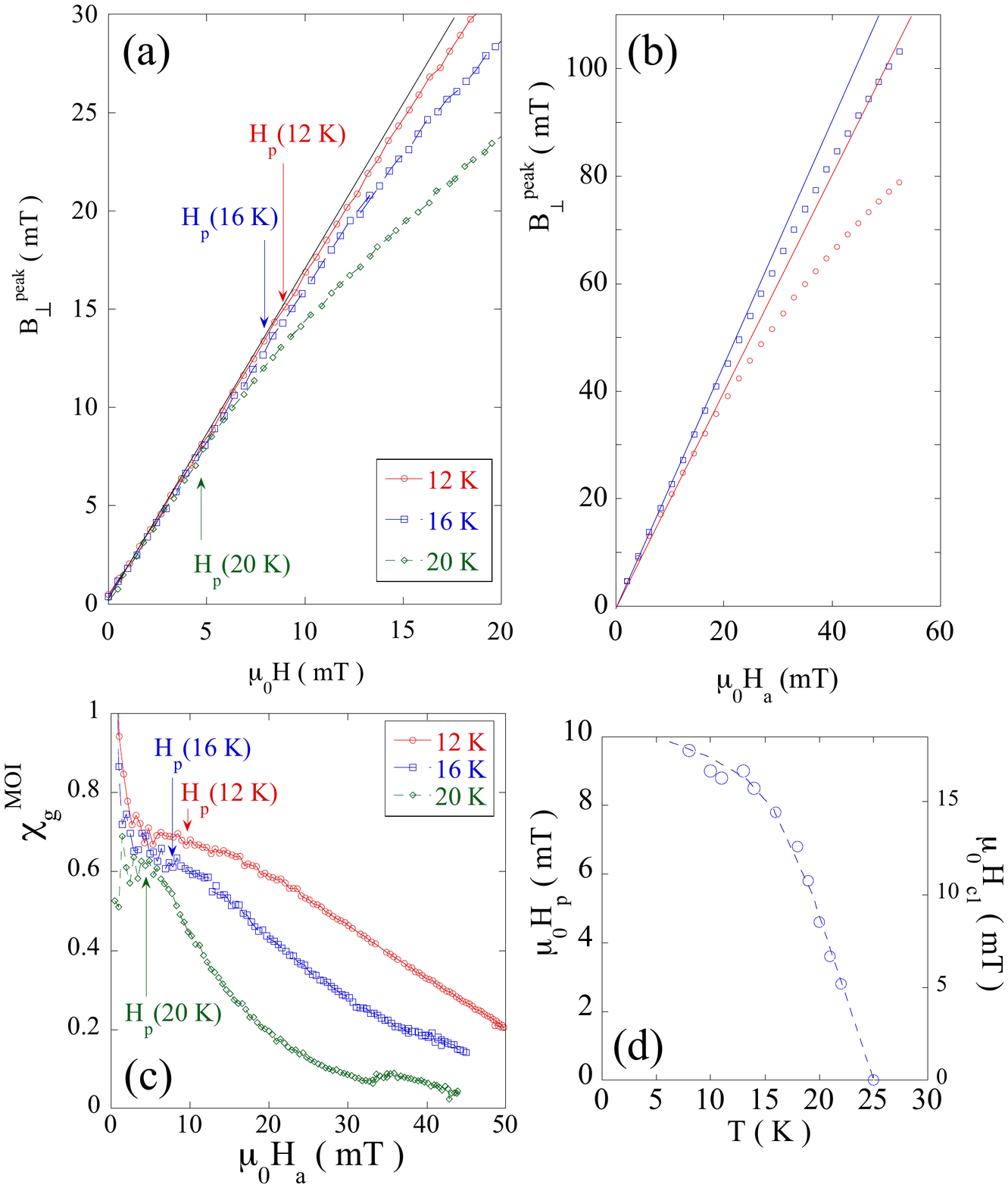}
\caption{\label{Fig_8}  Magnitude of the demagnetizing field peak at the edge of the Ba(Fe$_{0.925}$Co$_{0.075}$)$_{2}$As$_{2}$ crystal, measured with an aligned (a) and an oblique MO indicator (at 10~K, b). The upper and lower curves in (b) correspond to the right-hand and left-hand peak in Fig.~\protect\ref{Fig_4}. Panel (c) shows the geometrical susceptibility extracted from (a), and (d) shows the temperature dependence of the extracted penetration field $H_{p}$.}
\end{figure}
%
%%%%%%%%%%%% END FIGURE VIII %%%%%%%%%%%%%%%%
%

\subsection{Application to Ba(Fe$_{0.925}$Co$_{0.075}$)$_{2}$As$_{2}$ }
Fig.~\ref{Fig_8} summarizes the application of the above ideas to the  Ba(Fe$_{0.925}$Co$_{0.075}$)$_{2}$As$_{2}$ crystal. Panel (a) shows the evolution of the demagnetizing field maximum $B_{\perp}^{peak}$ for several temperatures; these curves allow for the extraction of the geometrical susceptibility in (c), which in turn indicates the effective probe-to-sample distance to be 12~$\mu$m -- rather larger than what is expected from the sole MO garnet thickness. However, applying the above mentioned correction for the in-plane field yields a more realistic distance of 9~$\mu$m, implying a gap of approximately 3~$\mu$m between the sample edge and the garnet surface. The temperature--dependent penetration field, extracted from the deviation from linearity, is shown in (d). Applying the aspect-ratio dependent relation between $H_{p}$ and $H_{c1} = \Phi_{0}/4\pi\mu_{0}\lambda^{2} \ln( \lambda / \xi )$ measured on samples of similar shape,\cite{Anna} one obtains the indicated $H_{c1}$--values, consistent with $\lambda( 5~{\mathrm K}) = 245$~nm\cite{LanLuan2011}  and a coherence length $\xi = 3.5$~nm (the flux quantum $\Phi_{0} = h/2e$).  

Drawbacks of the method include the need for a strictly rigorous calibration of the magnetic induction in order to obtain the correct curvature of the curves in Fig.~\ref{Fig_8}(a) and (c), and a high density of points in order to reliably extract $\chi_{\rm g}$. Nevertheless, the measurement at different locations on the sample boundary, or using an inclined MO indicator (Fig.~\ref{Fig_8}b) gives different slopes of the $B_{\perp}^{peak}(H_{a})$--curve and different $\chi_{\rm g}$, but the same penetration field $H_{p}$.   

\section{Conclusion}

In conclusion, the measurement of the applied field dependence of the demagnetizing field, and its expression in terms of a geometrical susceptibility, can be used to determine the sample-to-probe distance in magnetic imaging experiments on superconductors of finite thickness.  The measurement also offers an alternative means to determine the field of first flux penetration. 

A mathematical treatment of  full flux expulsion by the superconductor yields analytical expressions that allow one to describe the Meissner response of rectangular thick samples. Although not shown here, validation of the theory against finite element calculations was performed in the complete range of aspect ratios.

Anomalously low demagnetizing field peaks measured near the sample rims, and improbably large sample-to-probe distances such as these are obtained  from magneto-optical imaging experiments can be explained through the effect of the in-plane field component induced by the superconductor on the indicator garnet magnetization. Based on our calculations, we propose a straightforward method to correct for the in-plane field effect.

\section*{Acknowledgments}
A. Bad\'{\i}a acknowledges funding by the Spanish MINECO and the European FEDER Program (Projects MAT2011-22719, ENE2011-29741) and by Gobierno de Arag\' on (Research group T12). We are grateful to D. Colson and A. Forget of the CEA - IRAMIS - SPEC for providing the Ba(Fe$_{0.925}$Co$_{0.075}$)$_{2}$As$_{2}$ crystals. C.J. van der Beek acknowledges support from the MagCorPnic grant of the Triangle de la Physique du Plateau de Saclay.


\begin{thebibliography}{99}
\bibitem{ref4} Following Ref.\onlinecite{ref2}, one may define the function $f(s,m)\equiv E({\rm sin^{-1}}(s)\surd{m})-(1-m) F({\rm sin^{-1}}(s)\surd{m})$, where $m$ is inverted from $b/a=f(1,m)/f(1,1-m)$, and $s_u$ and $s_v$ from $u/a = f(s_u , 1-m)/f((1,1-m)$, $v/b = f(s_v , m)/f((1,m)$. $E$ and $F$ are incomplete elliptic integrals.
\bibitem{note} Namely, Eq.~(\ref{eq:intensity}) is obtained by simple substitution of Eq.~(\ref{eq:exp-real}) in the expression for the intensity that neglects the effect of the in--plane field, $I =  I_{0} \sin^2\left(  \alpha + {\mathcal V} M_{s} B_{\perp}^{MOI}/[B_{K}^{2}+B_{\perp}^{MOI 2}]^{1/2}\right) $.
\end{thebibliography}
\end{document}